\documentclass[preprint2]{aastex61}

\usepackage{amsmath,amssymb,amstext,mathtools}
\usepackage{graphicx}
\usepackage{epstopdf}
\def\tens#1{\ensuremath{\mathsf{#1}}}

\newcommand{\A}           {{\tens{A}}}
\newcommand{\M}           {{\tens{M}}}
\newcommand{\F}           {{\tens{F}}}

\newcommand{\E}           {{\tens{E}}}

\newcommand{\AWP}         {{\it A-Projection}}
\newcommand{\WBAWP}       {{\it WB A-Projection}}

\newcommand{\ostar}       {{\circledast}}

\received{March 3, 2017}
\revised{}
\accepted{}

\submitjournal{AJ}

\shorttitle{Direction Dependent Corrections in Polarimetric Radio Imaging I}
\shortauthors{Jagannathan~et~al.}

\begin{document}

\title{Direction Dependent Corrections in Polarimetric Radio Imaging I : \\
   Characterizing the effects of the primary beam on full Stokes imaging}
   
\correspondingauthor{Preshanth Jagannathan}
\email{pjaganna@nrao.edu}

\author{P.~Jagannathan}
\affiliation{National Radio Astronomy Observatory, Socorro, U.S.A}
\affiliation{Department of Astronomy, University of Cape Town, South Africa.}

\author{S.~Bhatnagar}
\affiliation{National Radio Astronomy Observatory, Socorro, U.S.A}

\author{U.~Rau}
\affiliation{National Radio Astronomy Observatory, Socorro, U.S.A}

\author{A.~R.~Taylor}
\affiliation{Department of Astronomy, University of Cape Town, South Africa.}
\affiliation{Inter-University Institute for Data Intensive Astronomy, and \\
Department of Physics and Astronomy, University of the Western Cape, Belville, South Africa}

\begin{abstract}
Next generation radio telescope arrays are being designed and
commissioned to accurately measure polarized intensity and rotation
measures across the entire sky through deep, wide-field radio
interferometric surveys. Radio interferometer dish antenna arrays are
affected by direction-dependent (DD) gains due to both instrumental
and atmospheric effects.  In this paper we demonstrate the effect of
DD errors for parabolic dish antenna array on the measured polarized
intensities of radio sources in interferometric images. We
characterize the extent of polarimetric image degradation due to the
DD gains through wide-band VLA simulations of representative point
source simulations of the radio sky at L-Band(1-2GHz). We show that at
the 0.5 gain level of the primary beam (PB) there is significant flux
leakage from Stokes $I$ to $Q$, $U$ amounting to 10\% of the total
intensity. We further demonstrate that while the instrumental response
averages down for observations over large parallactic angle intervals,
full-polarization DD correction is required to remove the effects of DD leakage. We also explore the effect of the DD beam on the Rotation Measure(RM) signals and show that while the
instrumental effect is primarily centered around 0 rad-m$^{-2}$, the effect is significant over a broad
range of RM requiring full polarization DD correction to accurately reconstruct RM synthesis signal.

\end{abstract}


\section{Introduction}

The next generation of radio polarimetric surveys are a part of a new era of wide-band, wide-area, full Stokes continuum surveys on parabolic antenna interferometer arrays in the US and on the Square Kilometre Array precursor telescopes in Australia and South Africa.

Such surveys will collect data over very large instantaneous bandwidths in the
GHz frequency regime and will require high-fidelity and high dynamic range
imaging in all polarization states of the incoming radiation over the 
full field-of-view (PB) of the array antennas.

The PB response of radio antennas varies with both direction
and frequency.  For altitude-azimuth mounted antennas the sky
brightness distribution rotates with respect to the antenna primary
beam as a function of the antenna parallactic angle. Consequently,
for long integration observations during which the parallactic angle
changes, the response of the array to a radio source includes an
instrumental component that varies with time, frequency and
polarization. These variations corrupt the wide-field image by
introducing image artifacts that are not removed by standard image
deconvolution approaches, thereby limiting dynamic range and fidelity.
The latter effect has been shown to be particularly important for the
polarization response \citep{2015ASPC..495..379J}.

The wide-band A-Projection (WB A-Projection or WB-AWP) algorithm  offers a solution to account for 
direction-dependent (DD) effects across a large observing bandwidth, and has been demonstrated to 
improve both dynamic range and image fidelity in wide-band total intensity images \cite{2013ApJ...770...91B}.
In this paper we examine the effects of wide-band DD errors on full-Stokes imaging performance based
on ray-trace models of the JVLA L-band full-Stokes PB response. 
In a subsequent paper we will assess the efficacy of the full-Stokes wide-band A-Projection algorithm in correcting for the DD effects. 
Using simulations of the sky brightness distribution we demonstrate the limits of imaging fidelity that 
can be achieved with classical calibration and imaging, and the levels at which the full-Stokes 
A-Projection algorithm becomes necessary to
enable high-fidelity and -dynamic full-Stokes
imaging of wide-band observations.

\subsection{Theory}
The measurement equation (ME) for a single interferometer, calibrated
for direction-independent (DI) terms\footnote{All terms that can
  be/are assumed to be constant across the field of view.}, is given by:
\begin{equation}
\label{eq:ME}
\vec{V}^{Obs}_{ij}(\nu,t)=W_{ij}(\nu,t)\int
\M_{ij}(s,\nu,t)\vec{I}(s,\nu)e^{\iota \vec{b}_{ij}.\vec{s}}d\vec{s}
\end{equation}
where $\vec{V}^{Obs}_{ij}$ is the visibility measured by a pair of antennas
$i$ and $j$, with a projected separation of $\vec{b}_{ij}$.
$W_{ij}$ are the effective weights, and $\vec{I}(s,\nu)$ is the sky brightness distribution 
 and is a function of direction $s$, and frequency $\nu$.  $\vec{I}(s,\nu)$ is a
 full-Stokes vector of images of the sky and $\vec{V}^{obs}$ is the
 vector of observed visibilities given by:
\begin{equation}
\label{eq:visvector}
\vec{V}^{obs}_{ij} = 
\begin{pmatrix}
V_{ij}^{pp} \\ V_{ij}^{pq} \\ V_{ij}^{qp} \\ V_{ij}^{qq}
\end{pmatrix}
~~~\&~~~
\vec{I} = \begin{pmatrix}
I^{pp}\\ I^{pq} \\ I^{qp}\\ I^{qq}
\end{pmatrix}\\
\end{equation}

\begin{figure}[ht!]
\includegraphics[width=\columnwidth]{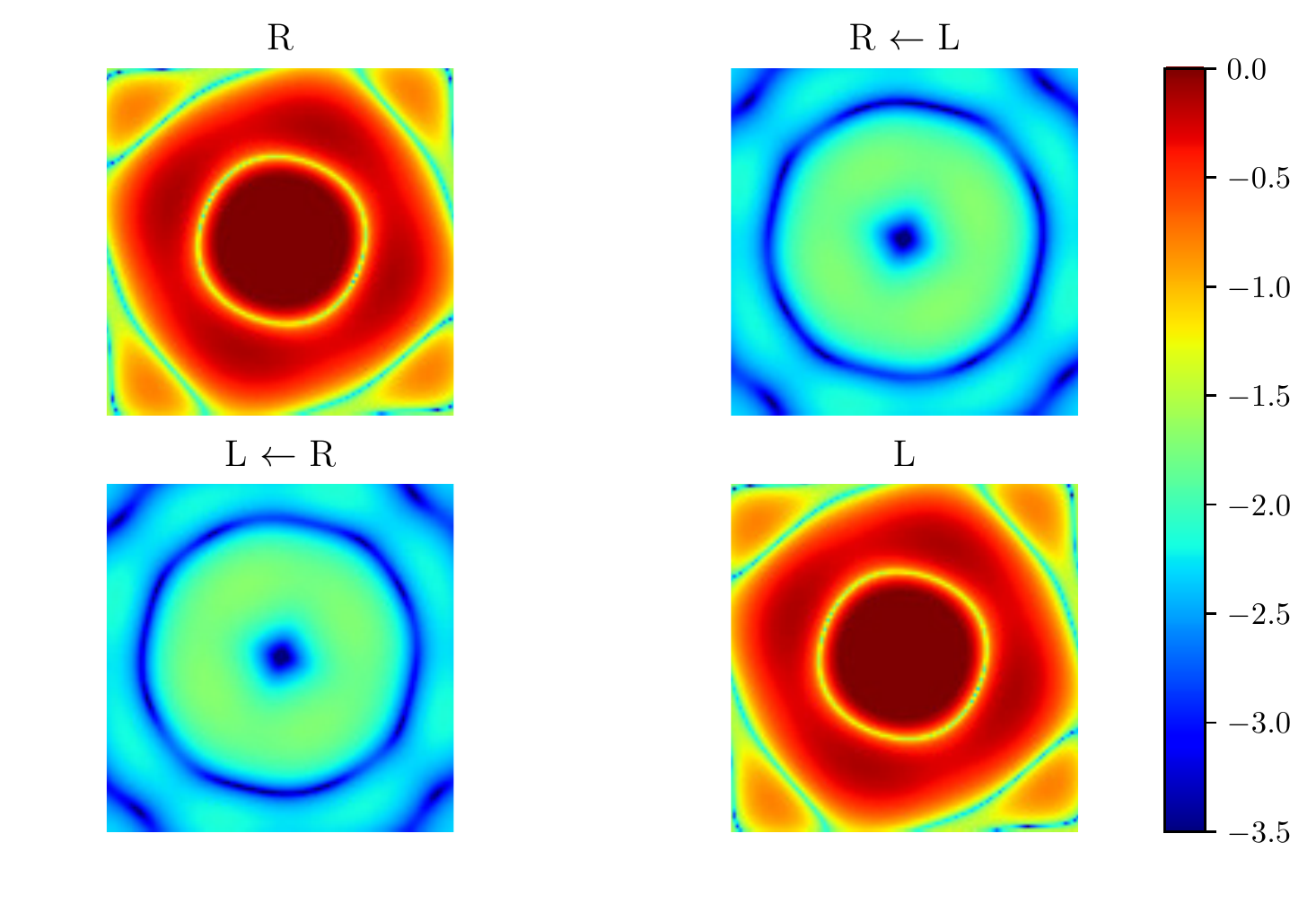}
\caption{The ray traced amplitude of the far-field complex  voltage pattern of the VLA Antenna, or the Fourier transform of the DD Jones Matrix, at $1.5$~GHz
The color scale is a logarithmic scale, and the plot has been normalized to unity at
the center of the {\tt R} and {\tt L} beams.
The direction of the arrows have 
been reversed from \citep{2015ASPC..495..379J} to be more in line with the vector 
notation in \citep{1996A&AS..117..137H}.}
\label{fig:Jones_amp}
\end{figure}

The super-scripts $p$ and $q$ represent the circular ($R$, $L$) or
linear ($X$, $Y$) polarizations states.  $\M_{ij}$, referred to as the
Direction-Dependent (DD) Mueller Matrix, encodes the mixing of the
various elements of $\vec{I}(s,\nu)$ in the measurement process.
$\M_{ij}$ can be written as an outer product of two antenna-based
Jones matrices as $\M_{ij}=\E_i\otimes \E_j^\dag$ where $\E_i$
describe the far-field electric field pattern for the antenna $i$.
Fig.~\ref{fig:Jones_amp} shows the amplitude of a complex-valued
$\E$-Jones matrix for a VLA antenna (in the circular polarization
basis) at L-Band computed using ray-tracing (for details, see \citet{2017paperII}).  
Similar to the DI G-Jones matrix, the diagonal elements of
$\E$-Jones matrix represent the DD complex voltage gain patterns and
the off-diagonal elements represent the DD leakage patterns in the
far-field.

Eq.~\ref{eq:ME} can be re-cast as:
\begin{equation}
\label{eq:ME2}
\vec{V}^{Obs}_{ij}(\nu,t)=W_{ij}(\nu,t)
\F\left[\left(\E_i(s,\nu,t)\otimes \E^\dag_j(s,\nu,t)\right) \vec{I}(s,\nu)\right]
\end{equation}
where $\F$ is the Fourier transform operator.
$\E_i$ and $\E_j$ both vary as a function of frequency, time (for
El-Az mount antennas) and direction in the sky.  The effects of $\E_i$
and the resulting mixing of polarization values encoded in $\M_{ij}$
are characterized in the sections below.  These effects due to
$\M_{ij}$ need to be calibrated and removed from $\vec{V}^{obs}$
during imaging for a precise measurement of the true sky brightness
distribution.  That this is optimally done as part of the imaging
process is discussed in \cite{2008A&A...487..419B}.

For the purpose of describing calibration of the observed data for
instrumental and atmospheric/ionospheric effects, it is more
convenient to write the right-hand side of Eq.~\ref{eq:ME} as:
\begin{equation}
\vec{V}^{Obs}_{ij}(\nu,t)=W_{ij}(\nu,t) \left[ \A_{ij}\star \vec{V}_{ij}\right]
\label{eq:RIME}
\end{equation}
where $\vec{V_{ij}} = \F I$, and
\begin{equation}
\A_{ij}=\A_i\ostar{\A_j}=\F \left[\E_i \otimes \E^{\dag}_j\right] 
=\F \left[\M_{ij}\right]
\end{equation}
$\A_i$ and $\A_j$ are the time- and frequency-dependent
 aperture illumination functions of the two antennas.
The diagonal terms of $\A_i$ represent the complex antenna gains for
the two polarizations while the off-diagonal terms represent the
polarization leakage across the antenna aperture. $\circledast$ is the outer convolution operator described in \cite{2013ApJ...770...91B}.
For identical antennas, the diagonal terms of $\A_{ij}$ are the Fourier
transforms of the {\it true} antenna far-field power pattern of the
four correlation products (PBs).

It is worthwhile to note here that $\A_i$ is the DD equivalent of the
DI G-Jones matrix in the \cite{1996A&AS..117..137H} formulation and
is the Fourier transform of the antenna far-field {\it voltage}
pattern.  Also note that $W_{ij}$ in the equation above is a scalar
and has no impact on the Full-polarization Wide-band A-Projection
imaging algorithm outlined below.  Therefore for brevity, we will
drop it from the equations, but note that as with other imaging
algorithms, the effects of various weighting schemes that modify
$W_{ij}$ are admissible.

Full-Polarization Wide-band A-Projection belongs to a
 general class of radio interferometric DD imaging algorithms that correct for the terms inside the
integral in Eq.~\ref{eq:ME}.  Briefly, this is done by applying the inverse of the terms
during convolutional gridding the imaging process
(transforming visibility data to the image domain).  These algorithms
however require a model for the antenna PB that includes all the
dominant effects that need to be calibrated. Given a model for the 
baseline aperture illumination, $\A_{ij}^M= \F \M_{ij}$, the \AWP\ algorithm
 computes the image as
$\F\left[\A^{M^\dag}_{ij}\star \vec{V}^{Obs}_{ij}\right]$ 
 and the resulting images are normalized by an appropriate function of
$\F\left[W_{ij}\left(\A_{ij}^{M^\dag}\star\A_{ij}\right)\right]$ (see
\cite{2013ApJ...770...91B} for details).

\begin{figure}[ht!]
\includegraphics[width=\columnwidth]{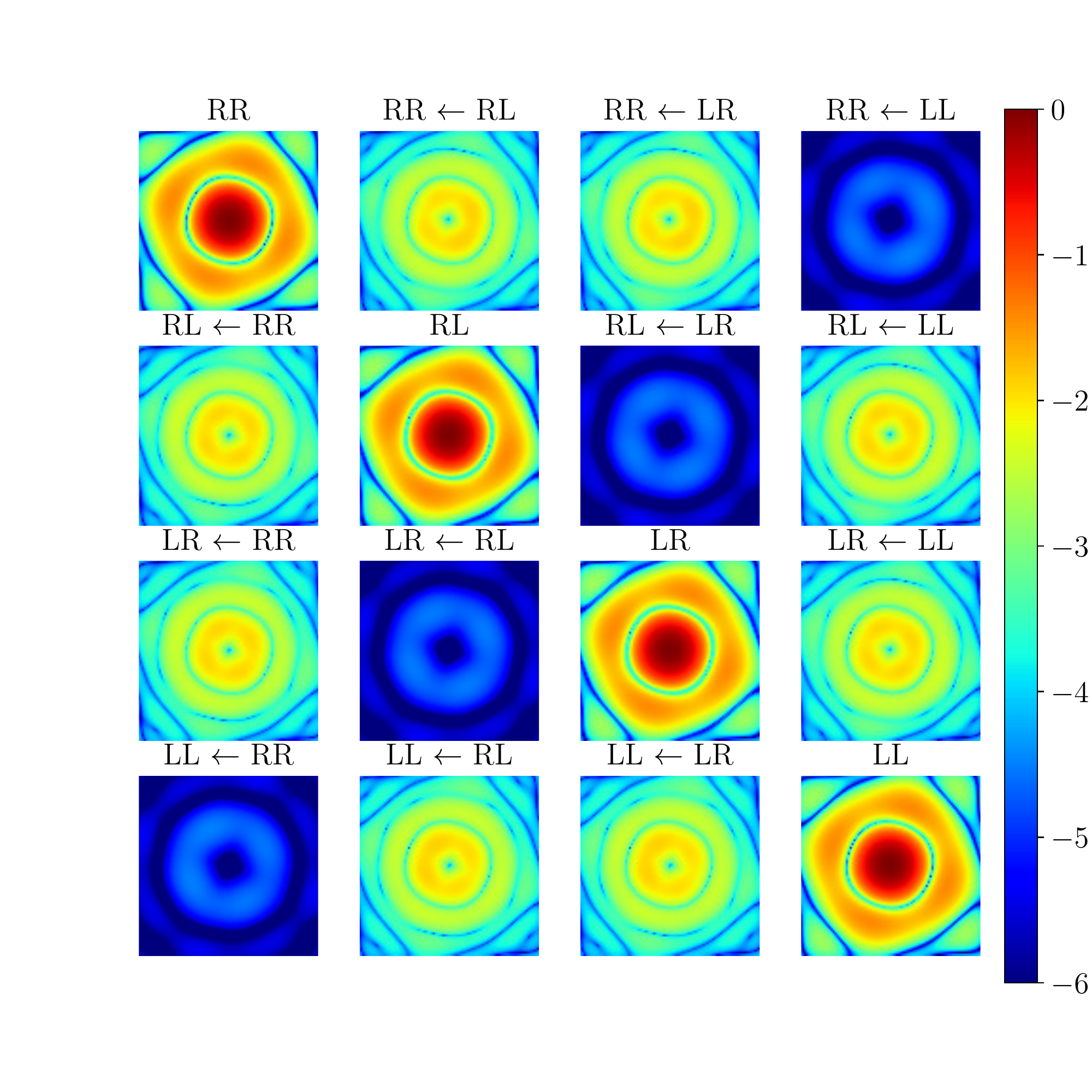}
\caption{Amplitude of the image domain complex Mueller Matrix for an arbitrary baseline $i-j$ at the VLA at $1.5$GHz. The color scale is a logarithmic scale spanning nine orders of magnitude.
The Mueller elements are computed using the normalized Jones matrix for the diagonal elements to have a peak value of unity along the pointing direction. The direction of the arrows have 
been reversed from \citep{2015ASPC..495..379J} to be more in line with the vector 
notation in \citep{1996A&AS..117..137H}.}
\label{fig:Mueller_amp}
\end{figure}

\begin{figure}[htbp!]
\begin{centering}
\includegraphics[width=\columnwidth]{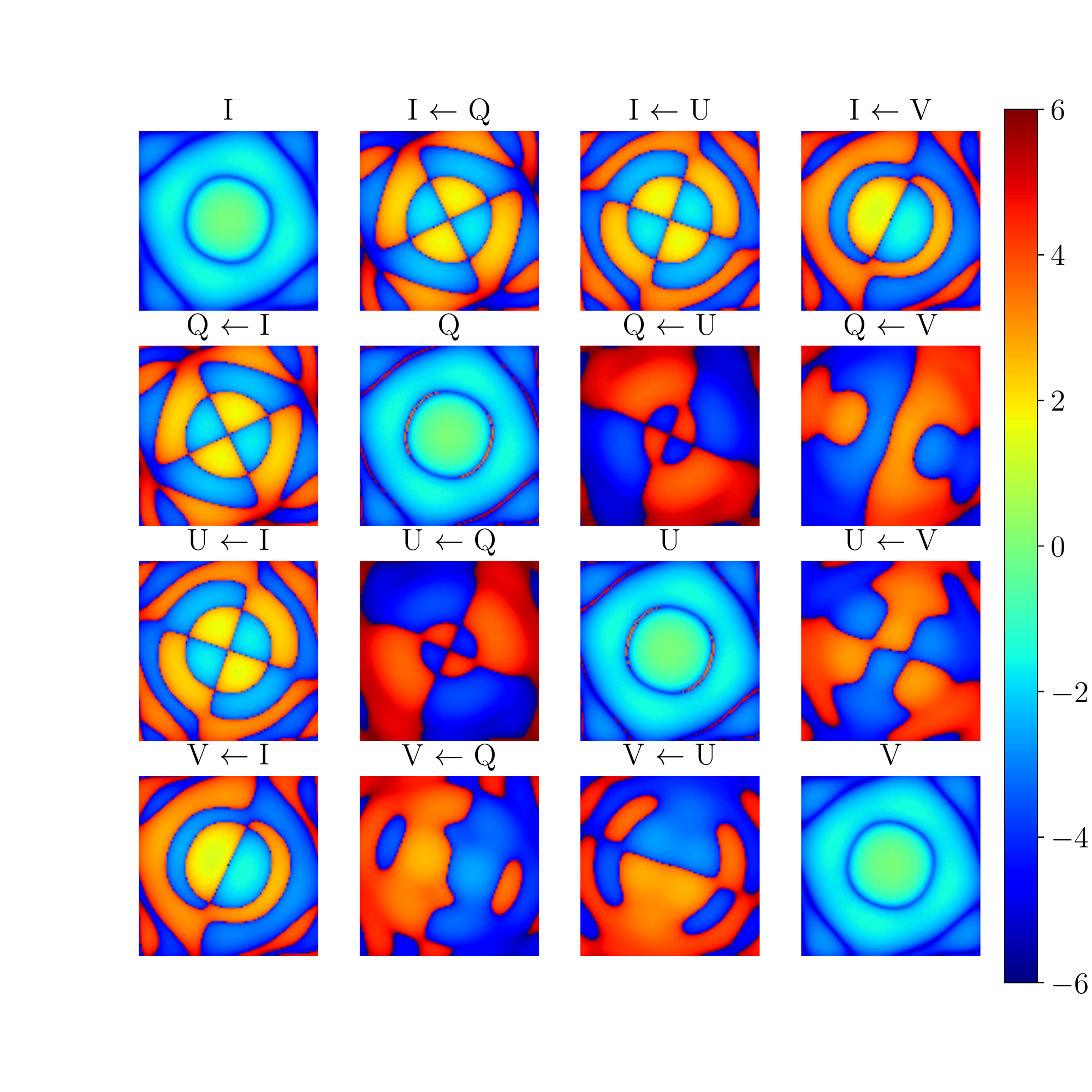}
\caption{The direction dependent Mueller in Stokes basis at $1.5$~GHz.  The images show
  $sign(M_{ij})\times \left(log_{10}(abs\left(M_{ij}\right)\right)$
  and the color scales is shown as the exponent of 10.  The leading
  order terms are along the diagonal.  The {\tt I<-V} and {\tt V<-I} term (due to
  polarization squint) are $\sim 2$ orders of magnitude lower than the diagonal. The
  {\tt I<-Q} and {\tt I<-U} terms (due to off-axis beam polarization)
  are $\sim 2.5 - 3$ orders of magnitude lower than the diagonal terms.}
\end{centering}
\label{fig:Mueller_Stokes}
\end{figure}

Fig.~\ref{fig:Mueller_amp} shows the $\M_{ij}$ matrix in circular
polarization basis for an EVLA antennas in L-band computed using
ray-tracing code (for details see, \citet{2017paperII}).  The diagonal elements of this
matrix are the antenna gain pattern for the sky signal in all
polarization products ($RR$, $RL$, $LR$,and $LL$) while the
off-diagonal elements encode direction-dependent mixing of the pure
polarization ($RR$ and $LL$) into the cross polarization ($RL$ and
$LR$).  $M_{ij}$ can be readily transformed into more familiar
Stokes-basis via a transform matrix
\begin{equation}
S = \frac{1}{2}
\begin{pmatrix}
1 & 0 & 0 & 1 \\
0 & 1 & 1 & 0 \\
0 & i & -i & 0 \\
1 & 0 & 0 & -1 
\end{pmatrix}
\end{equation}
Fig.~\ref{fig:Mueller_Stokes} shows a model for $\M_{ij}$ in the
Stokes basis, for EVLA antenna at L-Band.  The off-diagonal elements of
the first column shows the
familiar clover-leaf pattern for leakage of Stokes $I$ into Stokes $Q$,
and $U$.

\section{Polarization Effect of the antenna PB off-axis}
\label{sec:Point}

The continuous visibility full-polarization
vector-field $\vec{V_{ij}} = \F I$ is sampled by the interferometer via
$\A_{ij}$ as in Eq.~\ref{eq:RIME} to give $\vec{V}^{obs}_{ij}$.
$\A_{ij}$ terms introduce
the effects of the antenna far-field pattern in the observed data,
which if ignored, limits the imaging performance of the instrument
\citep{2016AJ....152..124R}. The $\A_{ij}$ varies with frequency,
time (due to relative rotation of the sky with for Alt-Az mount
antennas as, or as time dependent antenna pointing errors, structural
deformations, etc.) and polarization. The dominant variation of $\A_{ij}$ with
time for an Alt-Az mounted antenna is given by the change in parallactic angle
$\chi$, defined as the angle between
the source's hour circle and the great circle passing through the
source and the zenith as viewed by the antenna. 
For a source at declination $\delta$, the parallactic angle
changes with hour angle $h$, as 
\begin{equation}
\chi = \tan^{-1}\left \lgroup \frac{\cos L\cdot \sin h}{ \sin L\cdot \cos\delta-\cos L \cdot \sin\delta\cdot \cos h} \right \rgroup
\label{eq:parang}
\end{equation}
where, $L$ is the Geographical latitude of the antenna.  The WB-AWP
algorithm corrects for this parallactic angle-, direction-
and frequency-dependence of $\A$ for Stokes-I and -V imaging.

In the sections below we characterize the time and frequency dependent
errors introduced into Stokes $Q$ and $U$ for circularly polarized feeds
from an analysis including the effects of the off-diagonal terms of the Mueller matrix. 
We perform a suite of simulations that characterize the effects on a polarized
and unpolarized off-axis source.
We show that corrections for the time- and frequency-dependent PB in
all polarization including the effects of polarization leakages will be necessary (e.g. using a full-Stokes version
of the WB-AWP algorithm) for accurate reconstruction of the true
full-Stokes flux density vector in wide-field, wide-band imaging. In a
subsequent paper we will layout the framework and workings of the full-Stokes
WB-AWP algorithm that corrects for the entire Mueller
matrix during image reconstruction.

Linear polarized intensity can be expressed as 
\begin{equation}
 P = \Pi_{L} \,I\,e^{i\psi}
 \label{eq:pol}
\end{equation}
where $\Pi_{L}$ is the fractional linear polarization
and $\psi$ the electric vector polarization angle(EVPA). $\Pi_{L}$ and $\psi$ are defined in terms of the observed Stokes parameters as,
\begin{eqnarray}
 \Pi_{L}=\frac {\sqrt{Q^2 + U^2}}{I} \\
 \psi=\frac{1}{2}\, \tan^{-1}\left(\frac{U}{Q} \right)
\end{eqnarray}

The measurement equation in the image-plane can be cast in Stokes basis as,
\begin{equation}
\vec{I}^{M} = \sum_k \M^k \cdot \vec{I}
\label{eq:SIMEQ2}
\end{equation}
where $k$ is an index over time, frequency and baseline.
The vector $\vec{I}$ is the true incident full-Stokes sky brightness distribution.
 $\vec{I}^{M}$ is the measured apparent sky brightness
distribution. $\M^k$ encodes the response of the interferometer to the incident polarization vector. 
Expanding $\M^k$ in Eq.~\ref{eq:SIMEQ2} shows explicitly the dependence on the incident polarization.
\begin{equation}
\vec{I^{M}}=\sum_{k} \left\lgroup
 \begin{matrix}
 M_{\tt II}^{k} I + M_{\tt IQ}^{k} Q + M_{\tt IU}^{k} U + M_{\tt I V}^{k} V \\
 M_{\tt QI}^{k} I + M_{\tt QQ}^{k}  Q + M_{\tt QU}^{k}  U + M_{\tt QV}^{k}  V \\
 M_{\tt UI}^{k}  I + M_{\tt UQ}^{k}  Q + M_{\tt UU}^{k}  U + M_{\tt UV}^{k}  V \\
  M_{\tt VI}^{k}  I + M_{\tt VQ}^{k}  Q + M_{\tt VU}^{k}  U + M_{\tt VV}^{k}  V
 \end{matrix}\right\rgroup
 \label{eq:MuellerStokes}
\end{equation}
The diagonal of the Mueller matrix encodes the response of the 
interferometer pair to each individual Stokes parameter. 
The off-diagonal elements encode the leakage between Stokes parameters. 
$M_{IQ}^{k}$ and $M_{IU}^{k}$ encodes the amount of flux leaking from $I$ into $Q$ and $U$
respectively. 
Magnitude of leakage is $\sim$5$\times$10$^{-2}$ of $I$ for
the VLA at L-band at the half power point (Fig.~\ref{fig:Mueller_Stokes}), 
and grows with increasing distance from the beam center. 
Typical linear polarized intensity of astrophysical sources is at the
level of a few percent of Stokes-$I$ flux. Hence, the flux leakage
  results in a fractional error of up to 100\% in the wide-band
  full-Stokes measurements of typical astrophysical signals.
   It is noteworthy that the Mueller matrix element $M_{QI}^{k}$ 
representing the mutual coupling between Stokes $Q$ and $I$, has exactly the same magnitude
as $M_{IQ}^{k}$, within calibration errors. 
However, the amount of flux leakage is modulated by the intensity of the Stokes parameters. 
The amount of flux leaking from Stokes $Q$ into $I$ is $M_{IQ}^{k}\cdot Q$.
For a typical linear polarization of a few \%, the magnitude of flux leaking from Stokes Q into I is a 
$10^{-4}-10^{-5}$, which will be of a concern only for very high dynamic range imaging. 
However in cases of highly polarized emission such as may be seen in extended radio jets of some 
radio stars, leakage from $Q$ to $I$ of order $10^{-3}$ may occur in uncorrected images.

We can isolate the effects of leakage by recasting Eq.~\ref{eq:SIMEQ2} into a sum of the diagonal and off-diagonal elements,
\begin{equation}
 \vec{I}^{M}=\sum_{i,\nu,t} \left \lgroup M_{ii}\cdot I_i + \sum_{j,i\neq j} M_{ij}I_j \right\rgroup
 \label{eq:LeakageSum}
\end{equation}
The first term is the direction-dependent PB responses in
each Stokes.  The second term isolates the direction-dependent leakage
between Stokes flux leakage, which for linear polarization is
dominated by the direction-dependent Stokes I leakage.  Note that for
the time-variable component, all terms can be ignored and the
second term be represented as a constant only for very short
``snapshot'' observations with Alt-Az antennas.  This approach was
used for the NVSS \citep{1998AJ....115.1693C}. Observations with
  equatorial mount antennas or antennas with a third axis of motion to maintain
  a fixed parallactic angle \citet{2016PASA...33...42M} allow for 
  a simple correction in the form of a direction-dependent flux
  subtraction post imaging.
This technique was used to good effect for the Canadian Galactic Plane Survey \citep{2003AJ....125.3145T} 
using the equatorial mount antennas of the Dominion Radio Astronomy Observatory synthesis radio telescope.
For observations covering a large parallactic
range with Alt-Az antennas, these effects must be corrected as part of the deconvolution and imaging stage.

\begin{figure*}
\begin{centering}
\includegraphics[width=0.9\textwidth]{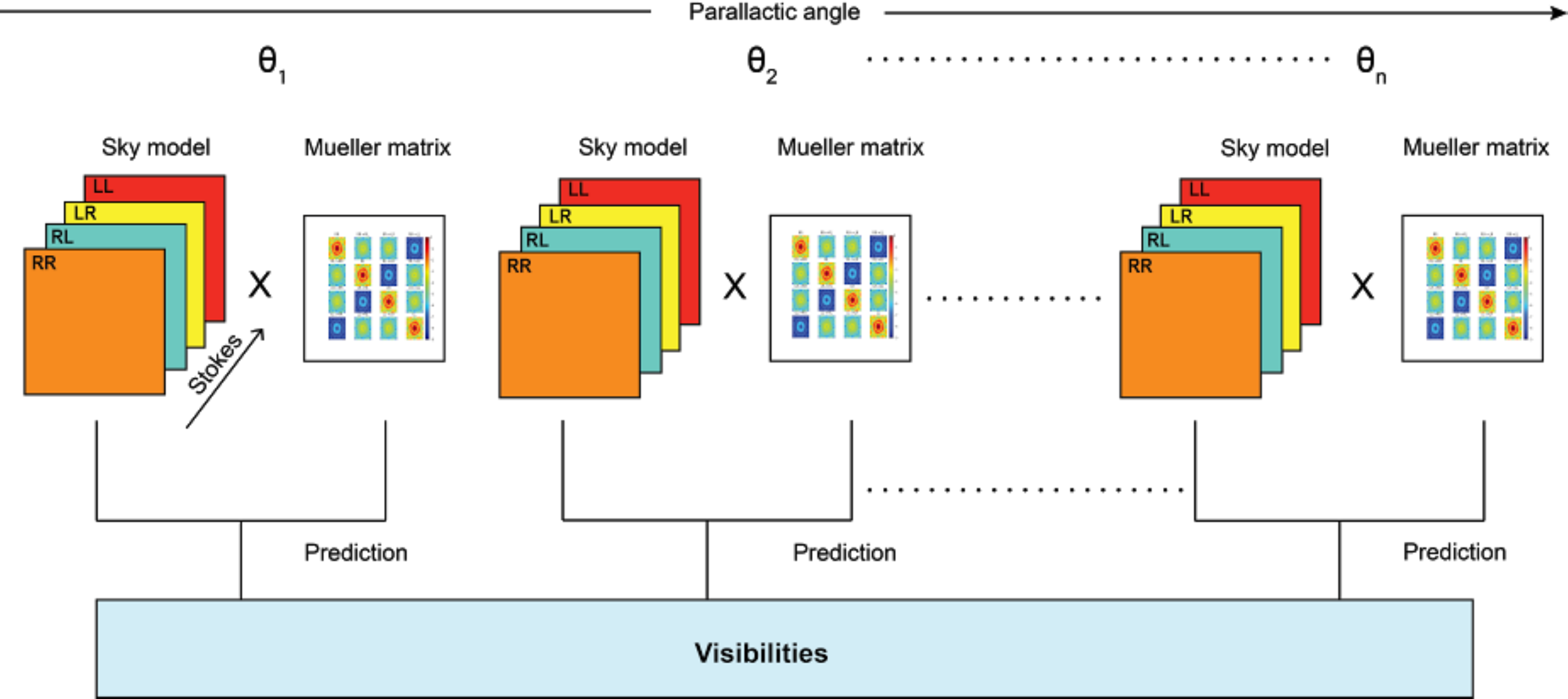}
\par\end{centering}
\caption{The imaging simulations followed the methodology outlined in the flow diagram. The sky model image was multiplied by the Fourier transform of the Mueller matrix for a change in parallactic angle of $5^{\circ}$}
\label{fig:Simulation} 
\end{figure*}

\section{Simulations}
 To examine the characteristics of the errors in polarized signals and their effect on astrophysical 
 measurements we carried out a suite of simulations. In each case we consider a simple point source
 located at an off-axis point in the PB.  This simple model allows for
 an intuitive understanding and interpretation of the resulting PB effects. 
 Three cases were considered. In the first
 case we consider the effect on an unpolarized point source.   We then consider two
 polarized sources, one with zero Faraday Rotation Measure (RM) and the second with a RM of
 100 rad\,m$^{-2}$.  These cases allow us to explore the effects of the frequency dependence of
 the Mueller terms on Faraday Rotation Measure synthesis.
 The simulations in all these cases utilized the ray-traced PBs using the 
implementation in the CASA R\&D
code base.
 
We simulated observations over an hour angle range of $-4$\,hr to $+4$\,hr and spanning 1\,GHz 
bandwidth in two frequency resolutions, the first with eight
$128$~MHz-wide spectral windows across the VLA L-Band(1-2\,GHz) to
understand continuum polarization imaging fidelity. The second
with 64 spectral windows with a bandwidth of 16~MHz each to understand the effects of antenna PB on Faraday rotation measure synthesis.
The sources were at Declination $+70^{\circ}$. 
An image domain sky model was created following Eq.~\ref{eq:SIMEQ2}. The sky model
was then used to
create visibilities given by the Eq.~\ref{eq:ME} for the VLA in C-Configuration. 
The multiplication of the Mueller Matrix with the sky models was carried out in a standalone 
python routine and the prediction from image plane to model VLA visibilities was carried out 
within the CASA framework. A block diagram of the simulation framework is provided in 
Fig.~\ref{fig:Simulation}.

\subsection{An Off-Axis Point Source as viewed by a single interferometric baseline}

We examine the effect of the PB off-axis through two simulations. The first is of an unpolarized point source of 1~Jy flux density located at the half power of the PB at the reference frequency of 1.5GHz. The second simulation is of a polarized point source at the same location with 1~Jy flux density in total intensity and 
a frequency-independent fractional linear polarization ($\Pi_L$) of 5\% with polarization position angle (EVPA) 
of 22.5 degrees.

\begin{figure}[ht!]
\includegraphics[width=9cm]{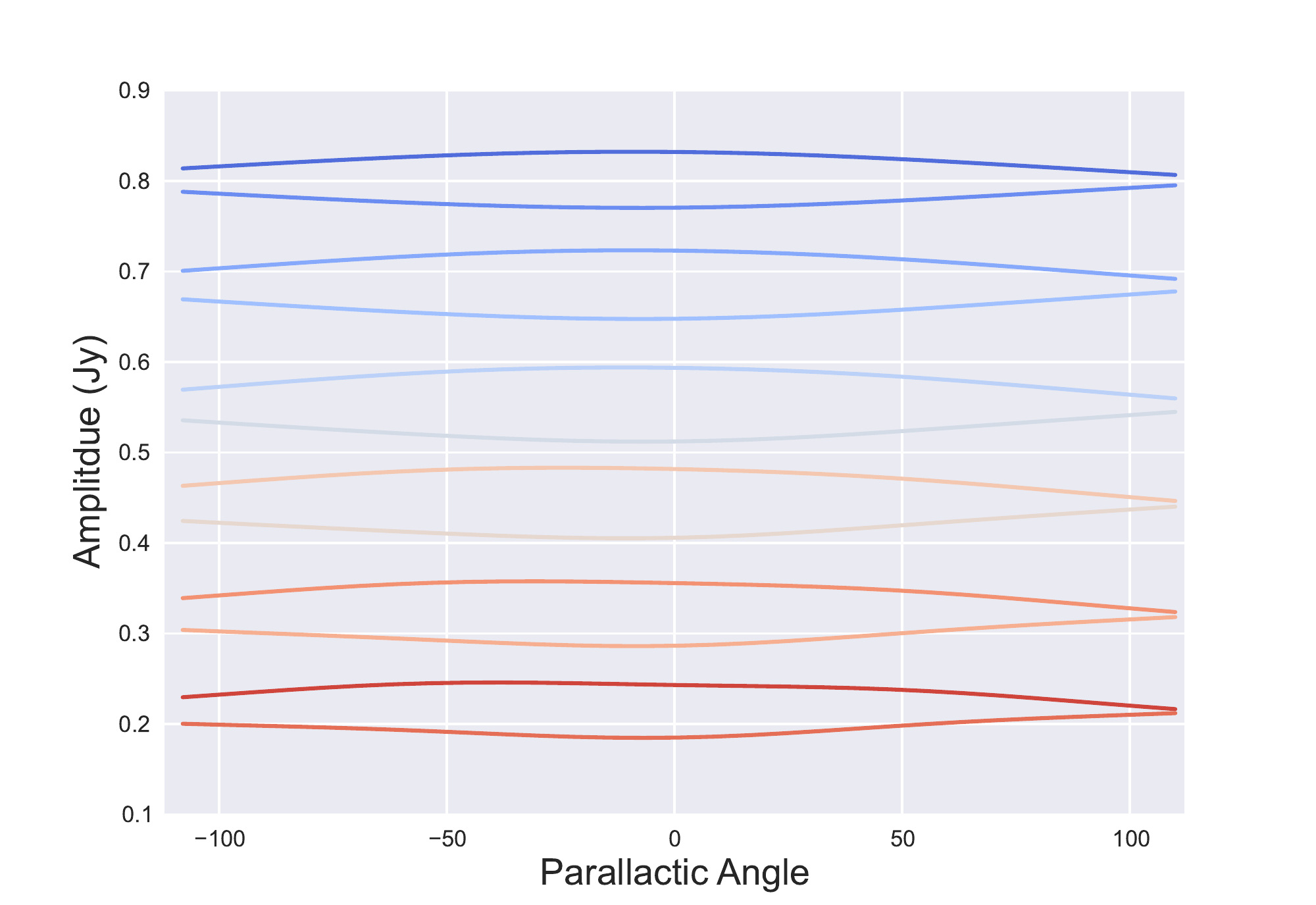}
\caption{The visibility amplitude of {\tt RR} (convex curves) and {\tt LL} (concave curves) of an unpolarized source situated at the half-power 
point at the reference frequency $1.5$GHz. The {\tt RR} and {\tt LL} show different amplitudes due to the squint of the VLA antennas. The color axis represents 6 spectral windows, spanning from 1\,GHz (blue) to 2\,GHz (red) in
steps of 0.2\,GHz}. 
\label{fig:RRLLbaseline}
\end{figure}

Fig~\ref{fig:RRLLbaseline} shows the {\tt RR} and {\tt LL} amplitudes for the unpolarized source at the half power point at 1.5\,GHz as a function of parallactic angle for a single baseline. The traces show data at six 
frequencies in steps of 0.2\,GHz from $1$GHz in blue to $2$GHz in red.
The differences in amplitude with frequency are due to the changing width of the PB across the large
fractional bandwidth. The source at 0.5 PB gain at 1.5\,GHz is at 0.8 PB gain at 1.0\,GHz and 0.2 PB gain
at 2.0\,GHz. This introduces a steep false spectral index that is corrected for in \WBAWP.
The similar but oppositely curved shapes of the {\tt RR} and {\tt LL} signals at each frequency arises from beam squint caused by the {\tt RR} and {\tt LL} beams having different pointing centers in the sky from the off-axis feed geometry of the VLA antennas. The effect of the squint is maximum at the half power of the PB and reduces in magnitude as the source moves away from the half-power point in either direction.

\begin{figure}[htbp!]
\includegraphics[width=9cm]{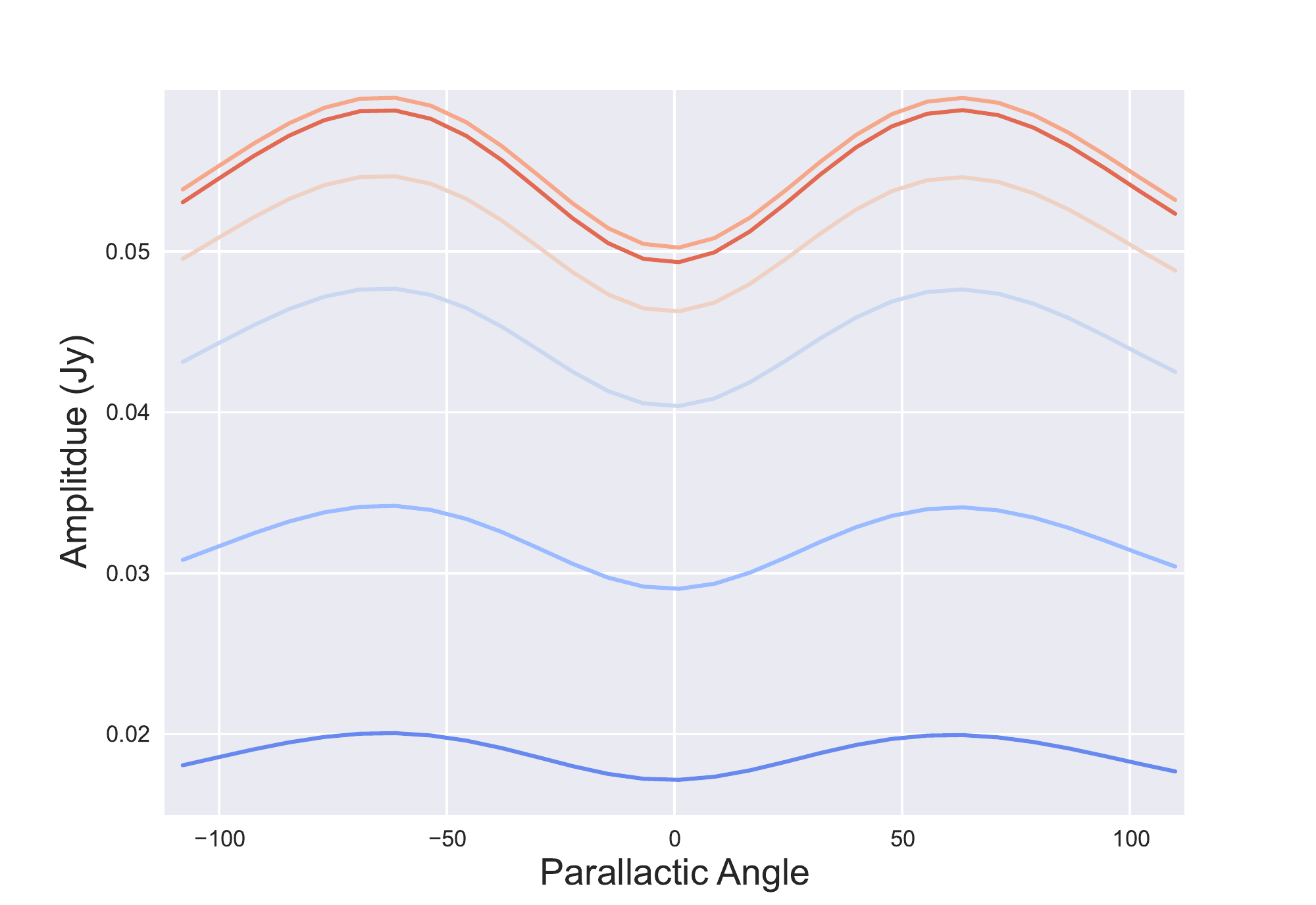}
\caption{Polarized flux density in {\tt RL} and {\tt LR} as a function of parallactic angle for the 
unpolarized source at the half-power point at 1.5\,GHz. Colors are as in Fig~\ref{fig:RRLLbaseline}. 
The signal arise from angle-dependent leakage from the source total intensity signal and also has
strong frequency dependence as the source moves out through the PB with increasing frequency.
{\tt RL} and {\tt LR} are both complex valued with
the same amplitude.  These curves therefore overlap exactly}.
\label{fig:UnPolLeakage}
\end{figure}

\begin{figure}[ht!]
\includegraphics[width=9cm]{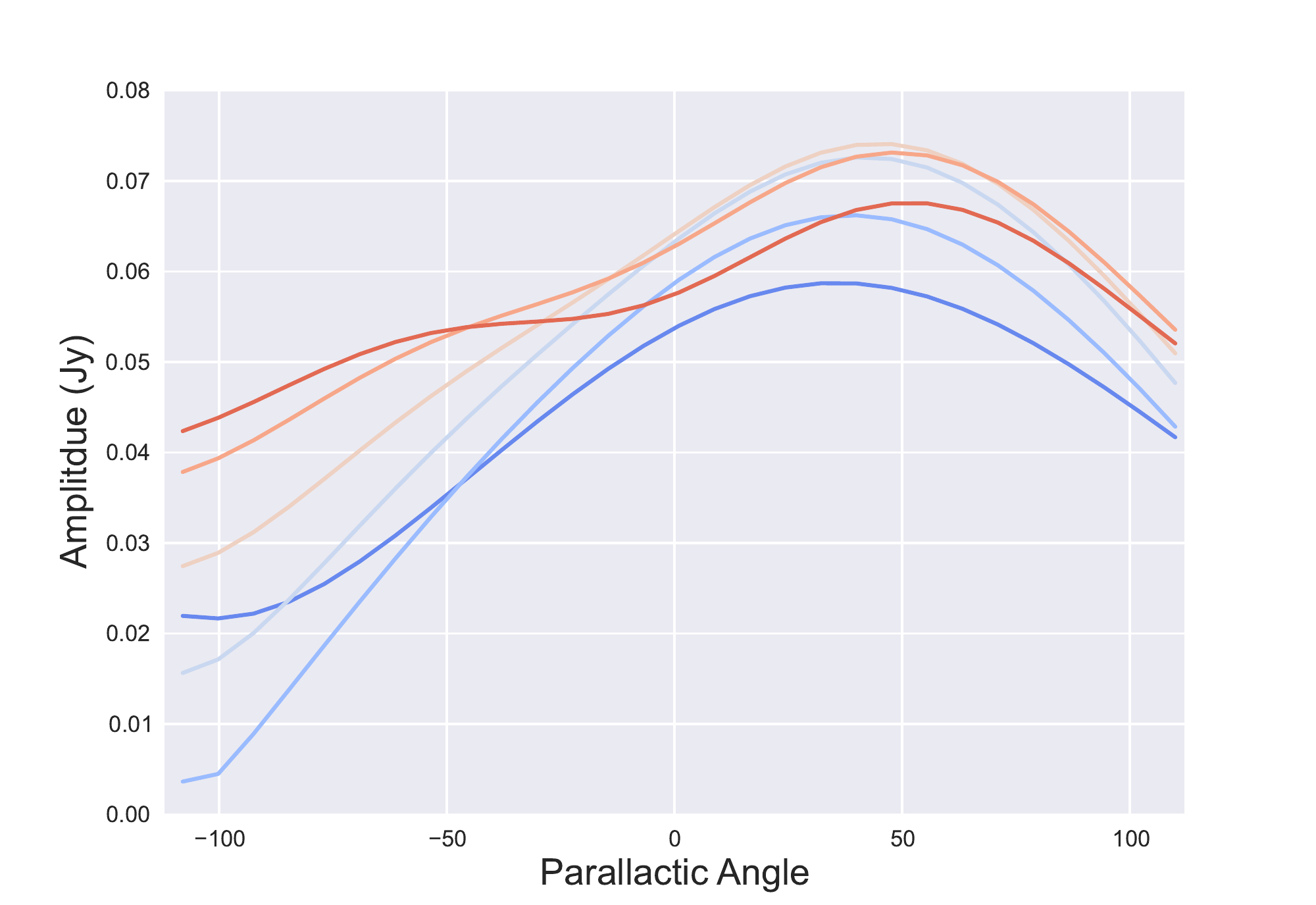}
\caption{Polarized flux in {\tt RL} and {\tt LR} as a function of parallactic angle for the 
5\% polarized source at the same location of the unpolarized source in Fig.\ref{fig:UnPolLeakage}. Color
are as in Fig.~\ref{fig:RRLLbaseline}. {\tt RL} and {\tt LR} are both complex valued with
the same amplitude.  These curves therefore overlap exactly}.
\label{fig:PolLeakage}
\end{figure}

Fig.~\ref{fig:UnPolLeakage} shows the amplitude of the cross-hand ({\tt RL}, {\tt LR}) visibilities 
for a baseline for the unpolarized source. 
The polarization signal arises from the leakage of flux from the total intensity of the
parallel hands {\tt RR} and {\tt LL} into the cross-hands, {\tt RL} and {\tt LR}. 
The leakage signal is a strong function of both parallactic angle and frequency.
Fig.~\ref{fig:PolLeakage} shows the same quantities for the 5\% polarized source placed at the same 
beam position. These traces show the combined effect of Stokes $I$ leakage and the other
Mueller matrix terms that are dependent on $Q$ and $U$.
The combined effects create a strong modulation of the polarized signal from less than 1\% to over 7\%. 

\subsection{Polarization Fidelity and Observing Interval}

\begin{figure}[ht!]
\includegraphics[width=9.5cm]{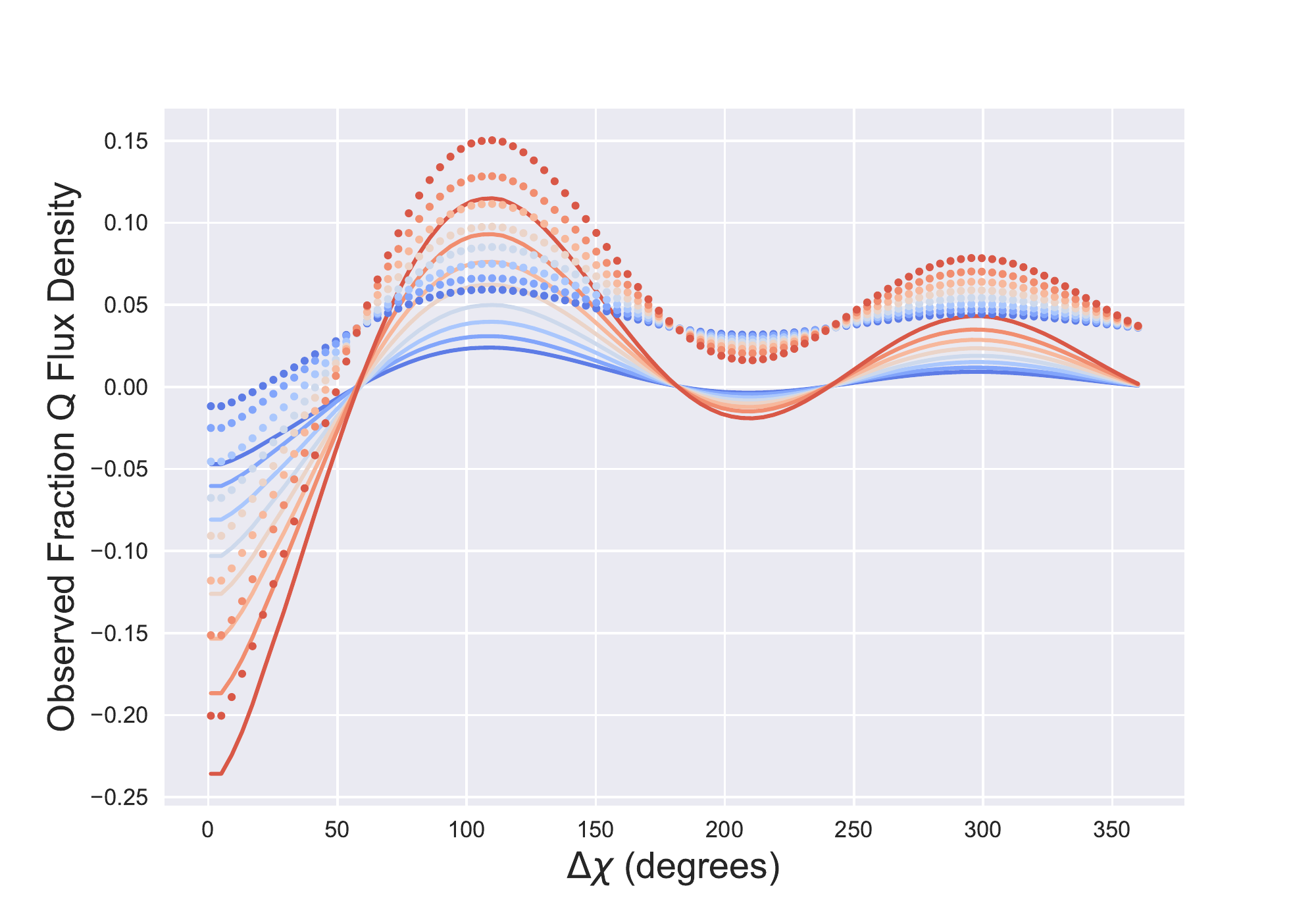}
\caption{The amplitude of fractional polarization ($Q^M/I^M$) as a function of the interval of parallactic angle during observations. The solid lines represent the unpolarized point source at the half power point(at 1.5 GHz). The dotted lines represent the  polarized point source of input Stokes $Q/I$ of 0.0353.}
\label{fig:Qleakage}
\end{figure}

\begin{figure}[ht!]
\includegraphics[width=9cm]{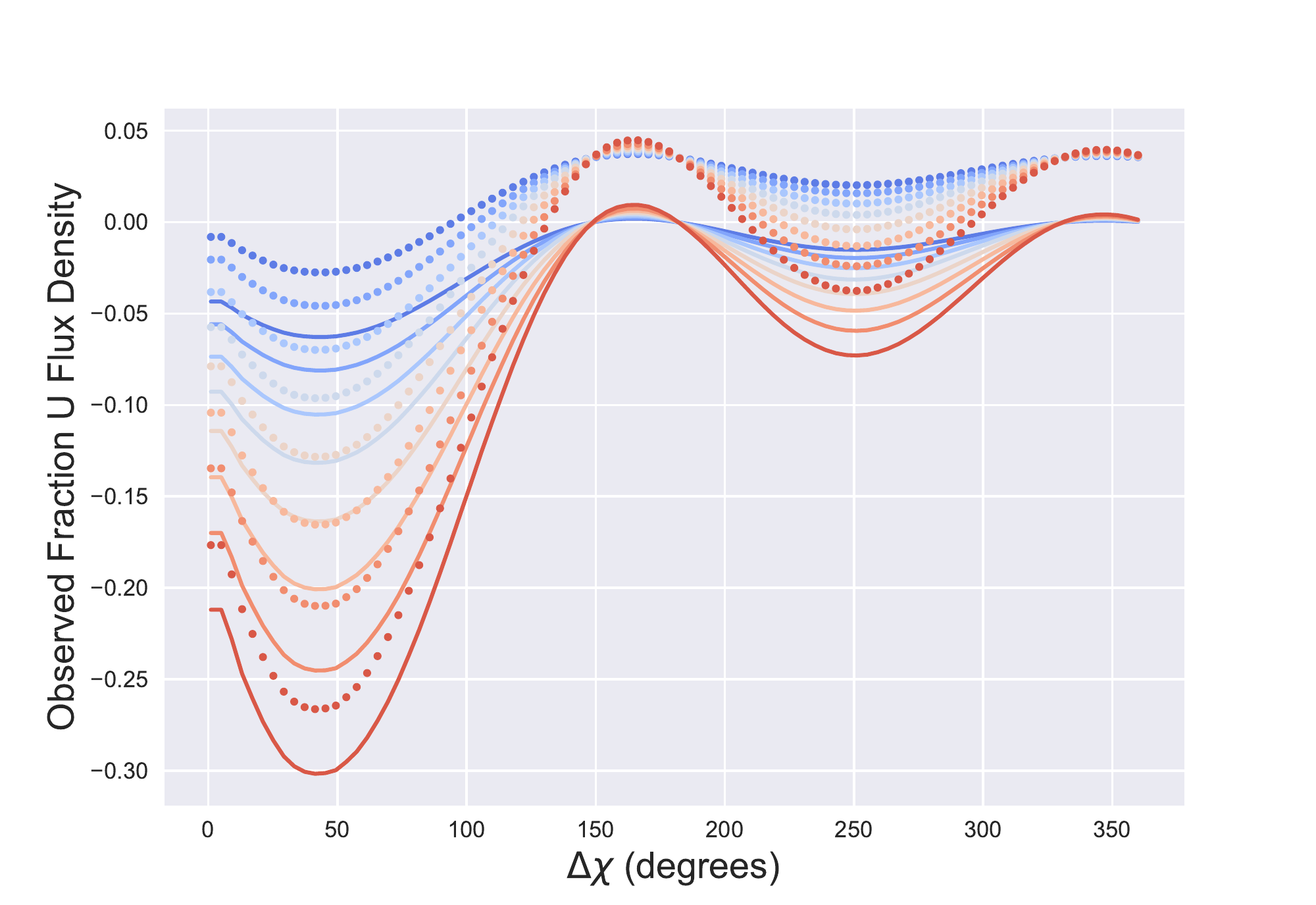}
\caption{As in Fig.~\ref{fig:Qleakage} but for $U^M/I^M$.}
\label{fig:Uleakage}
\end{figure}

\begin{figure}[ht!]
\includegraphics[width=9cm]{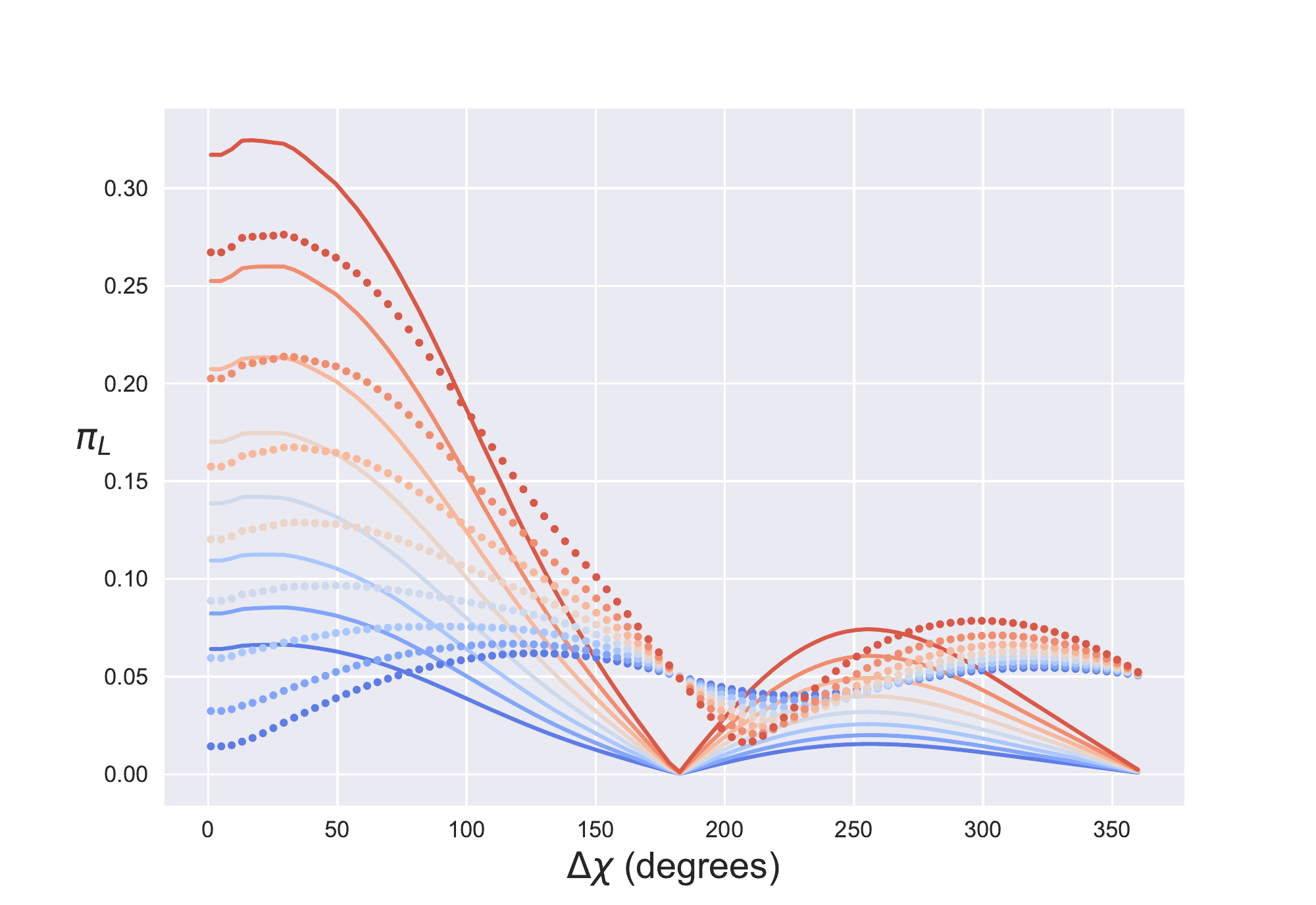}
\caption{As in Fig.~\ref{fig:Qleakage} but for $\Pi_L$}
\label{fig:Pleakage}
\end{figure}

The reconstructed flux density full-Stokes vector $\vec{I}^M$
during imaging results from a summation over time as given in
Eq.~\ref{eq:LeakageSum}. 
Over the course of an observation with total interval $\Delta t$, the parallactic angle will swing through
a corresponding interval in $\Delta \chi$. 
To show the effects of the duration of an observing on the error incurred in polarimetric flux density, we extended
the simulations over 24 hours and examine net polarization (equation.~\ref{eq:LeakageSum}) as a function of the parallactic
angle range $\Delta \chi$. 
In Figs.~\ref{fig:Qleakage},~\ref{fig:Uleakage},~\ref{fig:Pleakage} we show the observed fractional $Q$ and $U$ 
intensity ($Q^M/I^M$, $U^M/I^M$) and the total fractional polarization
$\Pi_L^M$ as a function of the parallactic angle interval $\Delta
\chi$, for both the unpolarized and polarized sources at the half-power point of the PB at 1.5\,GHz. 
In the plots the families of solid lines represent the response to the unpolarized source, while the dotted lines represent the polarized source with true $\Pi_L = 0.05$.  The colors represent frequencies as in the previous figures. 
For very small $\Delta \chi$, the error in fractional polarization can be very large.
At the high frequency end of the band the fraction leakage is $>0.25$ for $\Delta \chi < 50^{\circ}$, and
remains high until  $\Delta \chi$ approaches 180$^{\circ}$. At $\Delta \chi =180^{\circ}$ the net instrumental 
polarization sums to zero at all frequencies, and the true fractional polarization in all Stokes is retrieved. 
However, for typical observing intervals from a few minutes to a few hours, where the parallactic angle range
is much less than 180$^\circ$, 
instrumental polarization will be significantly larger than the true polarized signals over most of the PB. 

\subsection{Effect on Rotation Measure Synthesis}

At frequencies of a few GHz and below, the incident linear EVPA of a polarized source may be a complicated function 
of frequency due to Faraday effects either intrinsic to the source or due to propagation effects through
the intervening Galactic or extragalactic media. 
In the simplest scenario of an intervening Faraday screen the rotation of the EVPA is proportional to 
the square of the wavelength. 
\cite{1966MNRAS.133...67B} introduced the Faraday Dispersion Function (FDF) a measure of the polarization as a function of Faraday depth $\phi$ to measure the distribution of Faraday components effecting the polarization vector. 
This technique and associated analysis is becoming widely used as a powerful means to study the astrophysics of 
the Faraday effects in radio sources and intervening space.
\cite{2005A&A...441.1217B} introduced a window function and demonstrated that for a band-limited interferometric measurement the FDF can be approximated by 
\begin{equation}
\begin{split}
  \widetilde{F}(\phi) &\approx K\sum_{i=1}^{N} W_i P_i(\lambda^2)e^{-2i\phi(\lambda_{i}^2 - \lambda_{0}^{2})}  \\
 &\approx F(\phi)\star R(\phi)
\end{split}
 \label{eq:FDF}
\end{equation}
where $i$ is the channel of observation within a finite bandwidth,
$W_i$ is the weights per channel, $K$ the inverse of the
summed weights and $\lambda_{0}$ is the mean observing
wavelength. Eq.~\ref{eq:FDF} is a Fourier transform of the weighted complex polarized intensity. 
The observed FDF is a convolution of the true FDF with the Rotation Measure Spread Function (RMSF), or the Faraday depth point spread function given by
\begin{equation}
R(\phi) \approx K\sum_{i=1}^{N} W_i e^{-2i\phi(\lambda_{i}^2 - \lambda_{0}^{2})} 
 \label{eq:RMSF}
\end{equation}
The FWHM of the main lobe of the RMSF is a metric of the ``resolution" in Faraday space and is approximately
given by 
\begin{equation}
\Delta \phi \approx \frac{2 \sqrt{3}}{\Delta \lambda^2}\quad {\rm rad\,m}^{-2}
\end{equation}
with the wavelength $\lambda$ in meters.
For a band of spanning $1-2$\,GHz, $\Delta \phi \simeq 50$ rad\,m$^{-2}$.
The shape and range of the Faraday dispersion function of the instrumental leakage will be determined by 
$\Delta \phi$ and by the complexity of the wavelength
dependence of the time-averaged $M_{IQ}^{k}$ 
and $M_{IU}^{k}$ relative to $M_{II}^k$.  
A smooth variation with frequency will produce an instrumental signal largely confined to low values of RM.

\begin{figure}[ht!]
\includegraphics[width=9cm]{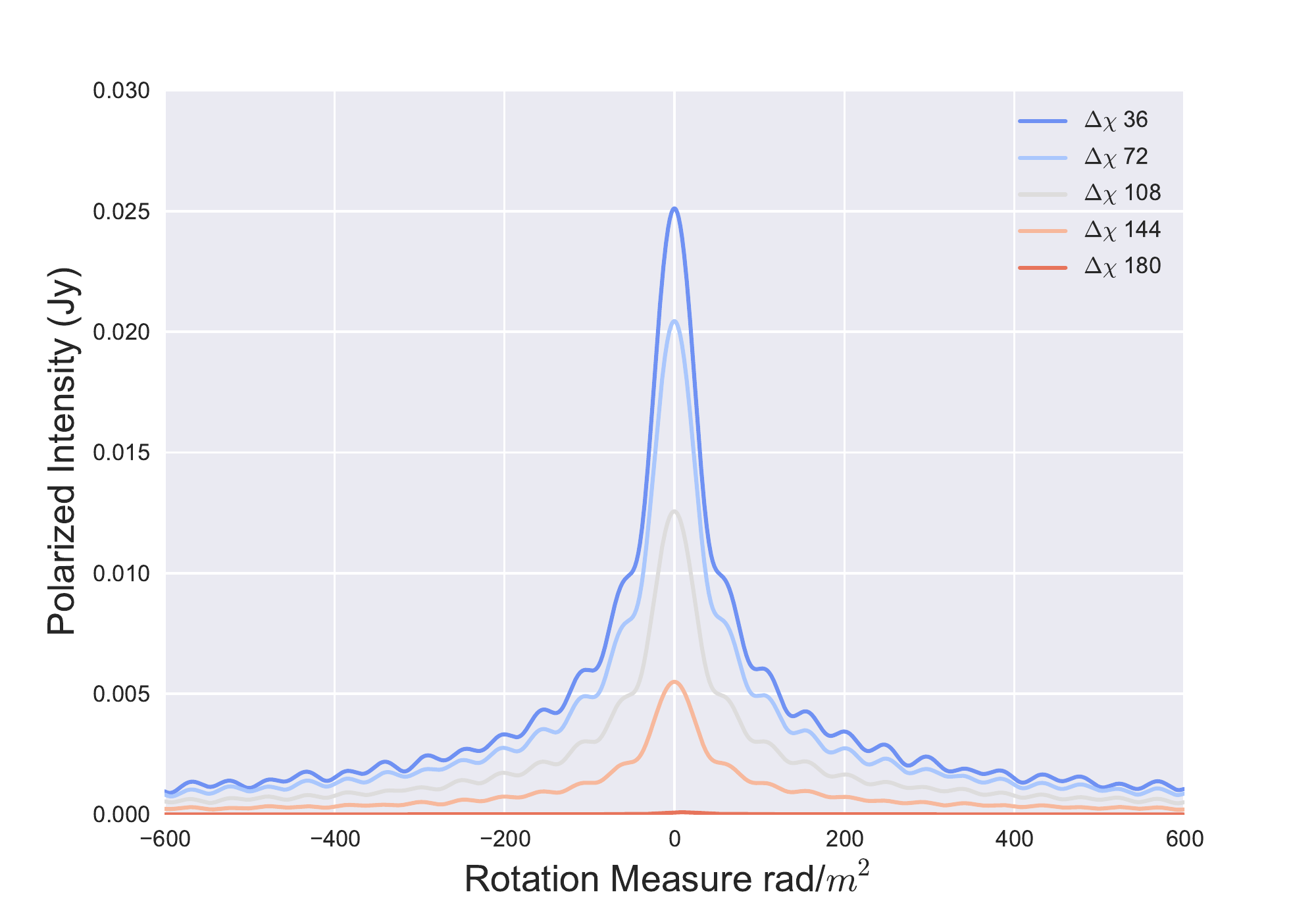}
\caption{The FDF of the instrumental leakage for the unpolarized source at the PB half power point at 1.5\,GHz.
In this plot the color now denotes increasing parallactic angle intervals spanned from 36$^\circ$ (blue) to 180$^\circ$
(red). The instrumental signal is maximum close to a Faraday depth of zero. The signal amplitude decreases for observations
with increasing $\Delta \chi$.}
\label{fig:polvsrmunpol}
\end{figure}

\begin{figure}[ht!]
\includegraphics[width=9cm]{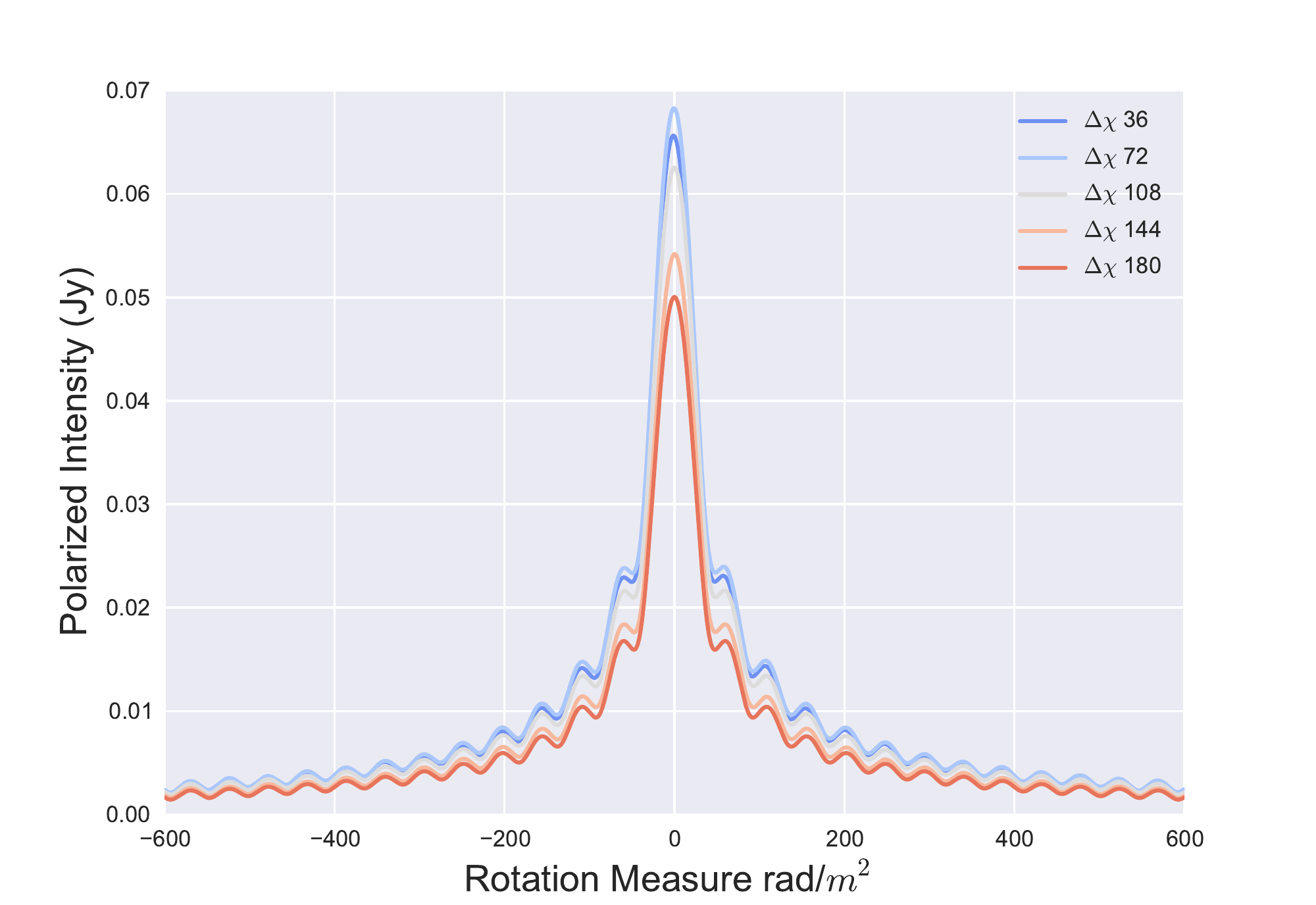}
\caption{Plotted is the linear polarized intensity of the polarized source ($\Pi_L$=0.05, $\psi=22.5$, RM=0 rad\,m$^{-2}$) 
at PB the power point at 1.5~GHz, as a function of rotation measure. The color axis denotes the different parallactic angle intervals spanned from 0 to 180 degrees. The spread in the RM around zero arises from the addition of the beam RM to the intrinsic source polarization.}
\label{fig:polvsrmpol}
\end{figure}

\begin{figure}[ht!]
\includegraphics[width=9cm]{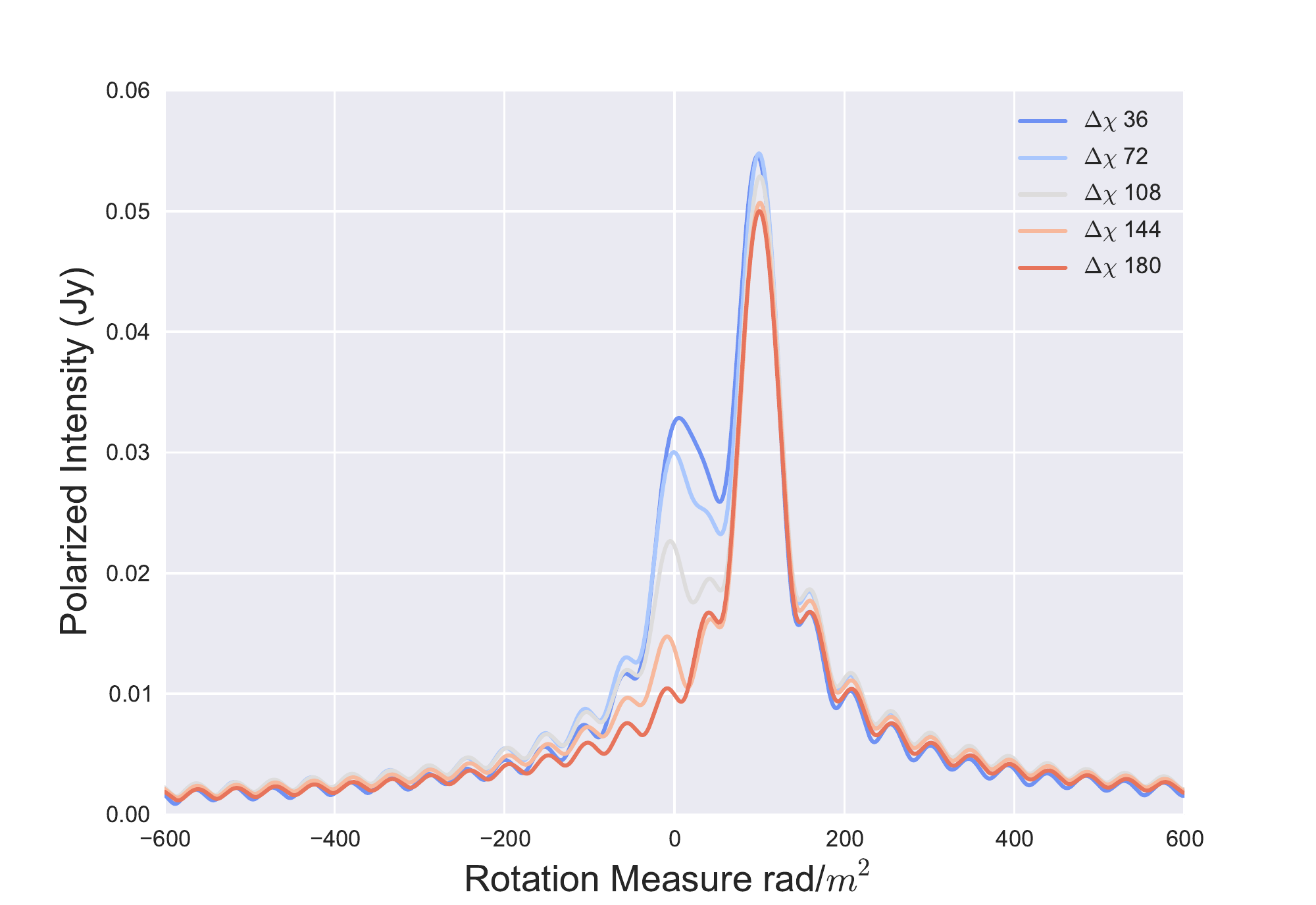}
\caption{Plotted is the linear polarized intensity of a polarized source (P=0.05, $\psi=22.5$,  RM=100) at half power at 1.5~GHz, as a function of rotation measure. The color axis denotes the different parallactic angle intervals spanned from 0 to 180 degrees.}
\label{fig:polvsrm}
\end{figure}

FDF of the 1-2 GHz off-axis response for the point source in our three simulations 
are plotted in Figs~\ref{fig:polvsrmunpol}, \ref{fig:polvsrmpol}, and \ref{fig:polvsrm}. 
For these plots the colors now indicate increasing observing intervals of the parallactic angle $\Delta \chi$,
ranging from $\Delta \chi = 36^\circ$ (blue) to $\Delta \chi = 180^\circ$ (red).

The FDF of the net Stokes $I$ leakage into $Q$ and $U$ from the unpolarized source 
at the half-power point of the beam is shown in Fig.~\ref{fig:polvsrmunpol}. 
The spectrum shows a central peak around RM = 0 rad\,m$^{-2}$. 
The``side-lobes" are largely from the RMTF determined by the overall bandwidth and the channelization.  
The amplitude of the central peak decreases with increasing $\Delta \chi$ as
the net instrument response reduces. The signal vanishes entirely at $\Delta \chi = 180^\circ$, as expected.
Fig~\ref{fig:polvsrmpol} shows the FDF of $P^M$ (the measured polarized intensity) of the source with $\Pi_L = 0.05$ and RM = 0 rad\,m$^{-2}$.
At small $\Delta \chi$ the true polarized signal is almost entirely removed.  The trace at $\Delta \chi = 36^\circ$
shows a very weak maximum shift slightly to negative RM. 
As $\Delta \chi$ get larger $P^M$ increases and approaches a better approximation of the true polarization, reaching
$P^M = P$ at $\Delta \chi = 180^\circ$.  This result demonstrates that for sources with low RM the instrumental 
polarization for observations with parallactic angle intervals less than $\sim$100$^\circ$ will significantly corrupts 
the true FDF signal. 

Fig.~\ref{fig:polvsrm} shows FDF for the case of a polarized source with RM of 100 rad\,m$^{-2}$. The signal
from the source is present at close to the correct amplitude and at the correct RM for all parallactic angle intervals. 
For shorter intervals ($\Delta \chi < 100^\circ$) there are strong spurious signals at RM=0 of similar amplitude to the source signal, and both the peak polarization and the RM of the peak of the signals from the polarized source differ 
from the true value by about 5\%.   As in the other cases, as $\Delta \chi$ approaches 180$^\circ$
the instrumental effect vanishes and the signal from the source approaches its true value. 

Faraday rotation synthesis thus offers the possibility to distinguish instrumental effects from true polarization
even in data uncorrected for off-axis effects for sources with large RM that are well separated in Faraday depth 
from the instrumental signal.  For the VLA at L-band RM significantly larger than about 50 rad\,m$^{-2}$ is 
required. While RM values of this magnitude are not uncommon in the plane of the Milky Way Galaxy due to the 
dense Galactic magneto-ionic medium \citep{2003ApJ...592L..29B},
at higher galactic latitudes such large RM values are the exception rather than the rule \citep{2009ApJ...702.1230T}. 
Moreover, broad-band polarimetry reveals complex FDF with signals over a range of RM in a significant fraction of sources, arising from internal Faraday effects \citep{2012MNRAS.421.3300O}.  Upcoming wide-area
surveys will require high precision polarimetry and Faraday Rotation synthesis over the full range
of RM.  Wide-band, off-axis polarization corrections therefore will be essential.

\subsection{Effects of Primary Beam Squash}

The effects of the antenna PB on polarimetric imaging described in the sections above is based on ray traced antenna models that account only for the leading order terms in phase (eg. \textit{squint}). Measured PB from holography \citep{2016VLAMEMO196} of the VLA antennas shows that the PB response is antenna dependent with different antennas displaying beam squash in addition to well known beam \textit{squint} of the VLA. While the PB \textit{squint} is a linear term in phase, which represents a displacement of the beam centers of {\tt R} and {\tt L} beams, PB \textit{squash} is a quadratic term in phase \citep{2001PASP..113.1247H}. \textit{Squash} is symmetric and is the superposition of \textit{defocus} and \textit{coma}, along with the addition of polarized emission from the sub-reflector feed legs. All of these effects alter the measured linear polarization response of the antenna. 

\begin{figure}[ht!]
\includegraphics[width=9cm]{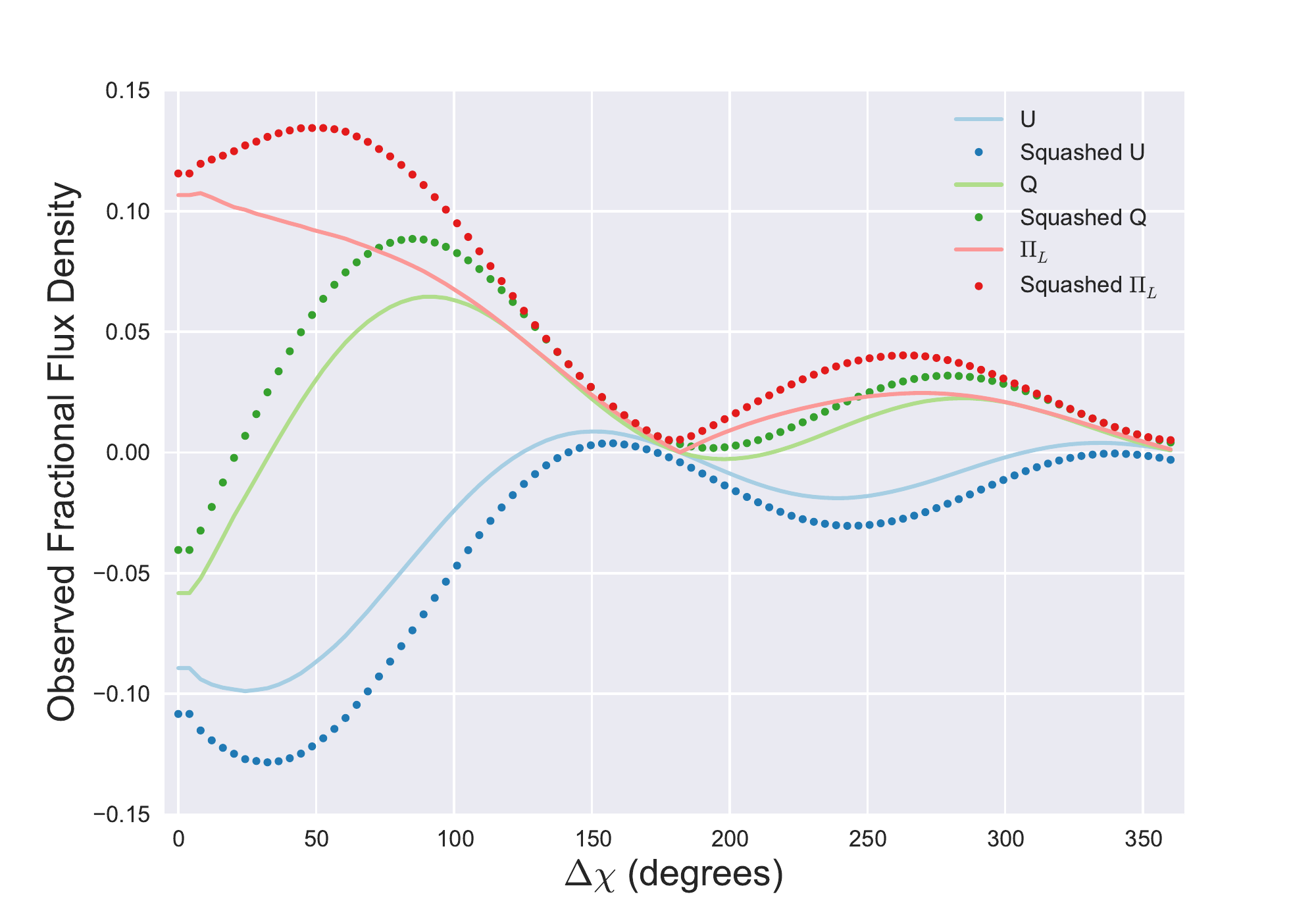}
\caption{Plotted is the leakage flux at the half power of PB at 1.5GHz, as a fraction of Stokes $I$ for different parallactic angle intervals $\Delta \chi$. The dotted(squashed) and solid(\textit{ideal}) in red represent leakage into to polarized intensity $\Pi_L$, the blue lines represent the Stokes {\tt U} and the green lines the Stokes $Q$.}
\label{fig:beam_squash}
\end{figure}

The most noticeable effect is the redistribution of power from the main lobe to the side-lobes and changes to the peaks of the two opposing polarization lobes in Stokes $Q$ and $U$. If the quadrapolar symmetry is altered, the flux leakage from Stokes $I$ leaking into the linear polarizations would not reduce to zero even when $\Delta \chi=180^{\circ}$. Fig.~\ref{fig:beam_squash}, shows the leakage as a fraction of Stokes $I$ for different intervals of parallactic angle, for an ideal beam and for a \textit{squashed} beam. The ideal antenna Stokes $Q$, $U$ and fractional polarization are plotted in solid lines in blue, green and red, while the squashed beam Stokes $Q$, $U$ are plotted in dotted blue, green and red respectively. The squashed beam shows higher levels of leakage for smaller parallactic angle intervals and does not average to zero $\Delta \chi=180^{\circ}$. The residual flux at $\Delta \chi=180^{\circ}$ is $\sim0.5 \%$ of Stokes $I$. 

The \textit{squashed} beam thus exacerbates the problem of leakage correction and
removes the symmetry that averages out of the leakage over
long integrations. The \AWP\ algorithm which can naturally
  account for the \textit{squash} term with an appropriate aperture models offers
the ideal solution for achieving noise-limited high fidelity imaging in polarization over the full PB.

\section{Conclusion}

The next generation of radio polarimetric surveys utilizing parabolic
dishes, spanning wide bandwidths and wide-area in all Stokes
parameters cannot achieve high fidelity in polarimetric imaging
using conventional imaging algorithms. The
effects of direction-dependent gain and polarization leakage 
in parabolic dishes with off-axis
  feed location is shown to create spurious polarized signals and significantly corrupt 
the polarization response at levels similar to or exceeding the true sky polarization.  
The off-axis effects from Stokes $I$ leakage can be ameliorated to some degree by observations over 
long averaging time intervals that provide a range in parallactic angle $\Delta \chi > 100^\circ$.
The net effect of the off-diagonal terms of the Mueller matrix averages to zero for
$\Delta \chi = 180^\circ$. However note that due to the time variability of the PSF, time averaging will lead to deconvolution errors during image reconstruction.
Observations that do not span such long time intervals or 
cover exactly symmetric hour angles about the meridian, the polarization leakage does not vanish. We also show that for realistic beam models with measurable second order errors (beam \textit{squash}) averaging of polarization leakage is not a viable solution.

The off-axis instrumental polarization response of the antenna in Faraday space is shown to be close to RM of 0 rad\,m$^{-2}$. 
For a typical RM signatures of astrophysical at GHz frequencies the off-axis leakage response alters the
Faraday dispersion spectrum, introducing spurious signals similar in scale to the sky polarization and 
corrupting the true sky RM signature.
Our simulations show that achieving polarimetric imaging simultaneously over wide-fields and wide-bands for
next generation deep and wide surveys will not be possible through conventional imaging methods that do not include DD corrections.

\acknowledgements
Support for this work was provided by the NSF through the Grote Reber Fellowship Program administered by Associated Universities, Inc./National Radio Astronomy Observatory. The National Radio Astronomy Observatory is a facility of the National Science Foundation operated under cooperative agreement by Associated Universities, Inc. Plotting in this paper has utilized the \textit{seaborn} package extensively. This work was done using the R\&D branch of the CASA code base.

\software{CASSBEAM, CASA, seaborn, matplotlib}

\end{document}